# TOWARDS EXPLANATION OF THE "INERTIA ANOMALIES" IN REALISTIC MEAN FIELD CALCULATIONS*

Nicolas Schunck and Jerzy Dudek

Université Louis Pasteur, Strasbourg I
*and*
Institut de Recherches Subatomiques
F-67037 Strasbourg, France

Careful theoretical studies of the nuclear collective inertia based on the non-relativistic mean-field theory and the cranking model (Ref. [1]) indicate that the moments of inertia of the super-deformed nuclei calculated using various standard methods and parameterizations exceed *systematically* the experimental values. Similarly, but for other reasons the relativistic cranking calculations based on the relativistic mean field method present, in our opinion, a different class of systematic deviations from experiment. The origin and possible explanations of these deviations are discussed briefly.

## 1. Introduction

The recent rapid progress in measurements of the rotational band properties of super-deformed nuclei has made it possible to learn about the nuclear theories in an efficient way, not only through the *similarities* between the experimental and the theoretical results, but also through the *systematic discrepancies* between them, Ref. [1]. The super-deformation studies enabled us to examine in particular the nuclear states with only weak, if not negligible, pairing correlations - although it is worth emphasizing that in many super-deformed nuclei the pairing correlations play certainly a non-negligible role.

The calculations of [1] have demonstrated, using both the realistic Woods-Saxon potential with the 'universal' parameterization and the self-consistent

---

Hartree-Fock approach with SkM* force, that in the discussed nuclei, the dynamical $\mathcal{J}^{(2)}$-moments (calculated without pairing) are *systematically larger* than the experimental values not only for the yrast super-deformed bands but also for the excited bands and this in many nuclei of the discussed mass range (even though only three nuclei were presented in Ref. [1] as an illustration). Most importantly, the introduction of the pairing correlations in these nuclei causes an *increase* in the $\mathcal{J}^{(2)}$-moments, a modification that goes precisely into the wrong direction[1].

Conversely, Relativistic Mean Field (RMF) calculations using various parameterizations of the effective interaction have implied much too low dynamical moments for the same nuclei. In [3], this effect was canceled by introducing a new term in the hamiltonian due to the broken time-reversal symmetry and interpreted in terms of the 'nuclear magnetism'.

In this paper, we present a simple phenomenological model based on the RMF approach that, we believe, is able to explain both anomalies by simple geometrical arguments and a non-optimal choice of the spin-orbit interaction in the case of the self-consistent RMF.

## 2. The Dirac Mean-Field Approach

The RMF approach combines the advantages of a fully microscopic treatment of the nucleus together with those of a relativistic approach (see [2] for a review). A wide variety of topics in nuclear structure has been studied by means of these methods and the reader is referred to the presentation of P. Ring in these Proceedings for the actual status.

In the case of the RMF theory the systematic discrepancies with respect to experiment exist already on the level of comparing the calculated single-particle energies with those of the doubly magic spherical nuclei. The single-particle level density there is known to be too low (the magic gaps systematically too large) and the often repeated argument says that this is because of the realistic effective mass used ($m^* \sim 70\%$ of $m_0$). It is our opinion that this argument must not be the only truth at least in the case of finite nuclei. First of all, many Skyrme Hartree-Fock approaches use regularly the effective mass that is numerically sometimes even lower - yet there are no really systematic and particularly strong discrepancies to be seen. Secondly, we found out that it is possible, by using the same form of the final Dirac equation as the one used within the RMF formalism, that the parameterizations of the effective mean-field potentials exists reproducing *simultaneously* the single particle level energies - both in terms of the level order *and* of the level density - and the overall geometrical properties: rms radii and charge distributions (cf. preliminary results in [5, 6] and [8]).

---

[1] It is worth emphasizing at this point that introduction of the pairing correlations causes at the same time a *decrease* in the $\mathcal{J}^{(1)}$-moments as it should.

The stationary RMF Dirac equation for the nucleons has the form

$$\left\{c\,\vec{\alpha}\cdot\hat{\vec{p}} + \hat{V}(\vec{r}) + \beta\left[m_0 c^2 + \hat{S}(\vec{r})\right]\right\}\psi_n = \mathcal{E}_n \psi_n \tag{1}$$

where $\{\vec{\alpha}, \beta\}$ are the usual Dirac 4×4 matrices, $m_0$ is the rest mass of the nucleon, $\psi_n$ are the eigen-functions and $\mathcal{E}_n$ the eigen-energies. Potentials $V(\vec{r})$ and $S(\vec{r})$ originate from the mechanism of exchange of the vector- and scalar-mesons, respectively. One often introduces the so-called effective-mass via [4]:

$$m^*(\vec{r}) = m_0 c^2 + \frac{1}{2}[S(\vec{r}) - V(\vec{r})]. \tag{2}$$

It is possible to show that around the Fermi level, the Dirac equation (1) may be expanded as a functional of $\varepsilon/2m^*(\vec{r})$ ($\varepsilon$ being the single-particle energy measured relative to the rest mass), within an error of less than 1%. This leads to the final Schrödinger-like equation for the "big component", $\xi_n$, of the Dirac bi-spinor[2]:

$$\left\{\frac{1}{2m^*(\vec{r})}\hat{\vec{p}}^2 + \hat{V}_{cen}(\vec{r}, \hat{\vec{p}}) + \hat{V}_p(\vec{r}, \hat{\vec{p}}) + \hat{V}_{so}(\vec{r}, \hat{\vec{p}}, \hat{\vec{s}})\right\}\xi_n = \varepsilon_n \xi_n. \tag{3}$$

The effective mass, the spin-orbit and the linear-momentum potentials, $m^*(\vec{r})$, $\hat{V}_{so}(\vec{r}, \hat{\vec{p}}, \hat{\vec{s}})$ and $\hat{V}_{\vec{p}}(\vec{r}, \hat{\vec{p}})$, respectively, depend only on the difference of $\hat{V}(\vec{r})$ and $\hat{S}(\vec{r})$, while the central potential is a sum of the two. Both these functions, i.e. the sum and the difference, are replaced by the Woods-Saxon forms each of which depending on the radius-, diffuseness- and depth-parameters (cf. e.g. [5]). We fit these parameters to the experimental results on the single-particle energies and of the r.m.s. radii of eight spherical doubly-magic nuclei by using a specially designed semi-automatic fitting program. From the parameters obtained for each doubly-magic nucleus, it is also possible to extract their systematic dependence on the isospin and on the nuclear mass. Such a relation allows to get an approximated set of parameters for any nucleus in the nuclear chart. The complete results of this fitting algorithm will be presented elsewhere [8].

Two comments are in place here. Firstly, the new fits guarantee that not only the positions of the levels close to the Fermi level are well reproduced but also that the deeply bound states (the lowest $1s_{1/2}$ states) are close to their experimental positions; those are known experimentally in a few cases. Secondly, of course the geometrical features (rms radii) are reproduced as well.

---

[2] An analog equation is obtained for the 'small component' of the Dirac bi-spinor. Let us notice that the eigen-problems for the big and the small components separately are not independent so that solving Eq. (3) is in fact *equivalent* to solving Eq. (1) within the effective-mass approximation. The reader is referred to [5] for comments.

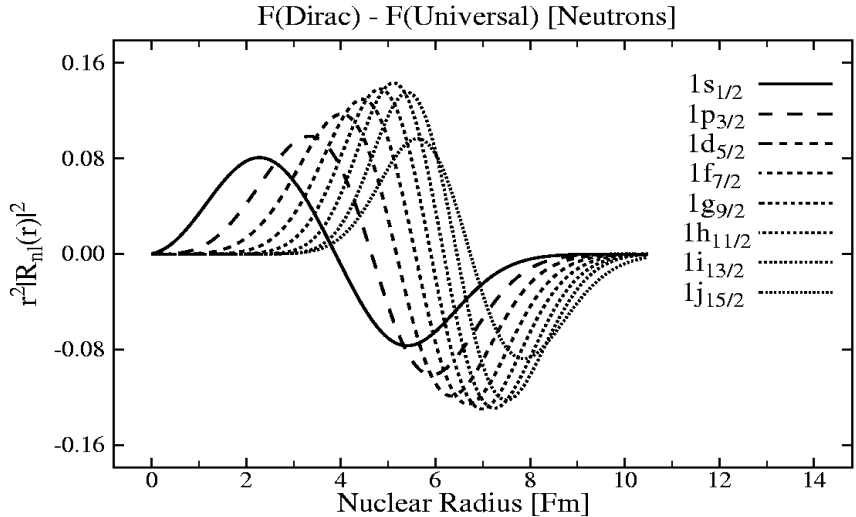

Fig. 1. *Differences between the nucleonic probability distributions for the orbitals indicated obtained by using the Dirac Woods-Saxon parameters of this work and the 'universal' Woods-Saxon parameters.*

The fact that the experimental positions of the deeply bound states are reproduced at the constant r.m.s. radii has a direct influence on the single nucleonic wave functions and thus on the nuclear mass distribution. Figure 1 shows that the nucleonic mass is distributed much closer to the nuclear center and we have demonstrated in [8] that this mechanism explains the anomaly obtained in [1].

The most interesting features come from the comparison of the results in our approach with those obtained in the RMF theory. For this purpose we have used the equivalent Woods-Saxon potential parameters that have been fitted in [7]. Solving Eq. (3) with these parameters is a good approximation to the RMF self-consistent results. In Figure 2, the $\mathcal{J}^{(2)}$-moments in $^{152}$Dy for our parameterization of the relativistic mean-field and for the RMF-equivalent Woods-Saxon parameterization are shown. It is worth noticing that: (a) the configurations are the same in both cases (in particular the number of intruders), (b) the 20-30 % discrepancy between theory and experiment stressed in [4] and visible in the RMF-equivalent parameterization is not present with our parameterization. It is to be emphasized that our parameterization assures the simultaneous reproduction of several experimental features like the single-particle level order *and* the level density, as well as the dynamical moments and relative alignments in the nuclei from our test region around $^{152}$Dy (the remaining a few percent discrepancy is attributed to pairing, without additional parameter fit). The origin of the differences between the Dirac Mean-Field and the RMF equivalent hamiltonian lies in the spin-orbit term as discussed in detail in [8].

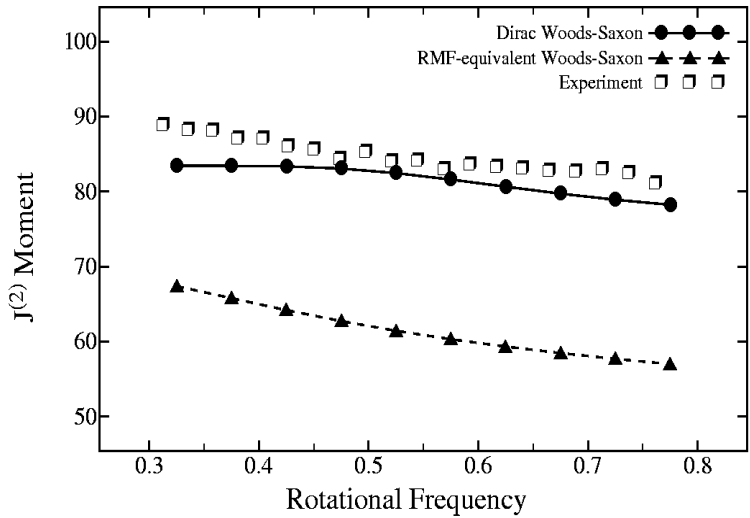

Fig. 2. $\mathcal{J}^{(2)}$-moment in $^{152}Dy$ in the Dirac Mean-Field approach (full circles full lines) and in the RMF-equivalent parameterization (full triangles dashed line).

In this paper, we suggest that the geometry of the mean-field potential does play a crucial role in the accurate calculation of high-spin features, and in particular, that the incorrect geometry (here: not deep enough effective central potentials) could be a possible explanation of the systematic discrepancies for the $\mathcal{J}^{(2)}$-moments observed in nuclei around $^{152}$Dy. In the RMF theory, a different type anomaly originates from a non optimal spin-orbit potential. More details will be presented in a forthcoming paper [8].